\RequirePackage{fix-cm}
\documentclass[smallextended]{svjour3}       
\smartqed  
\usepackage{graphicx}
\usepackage{natbib}
\usepackage{color}
\usepackage{amssymb}
\usepackage{lineno}
\usepackage{txfonts}
\usepackage{amstext}
\usepackage{mathabx}
\usepackage{multirow}
\usepackage{ulem}
\usepackage{lscape}
\usepackage{txfonts}
\usepackage{subfigure}
\usepackage{float}

\usepackage{hyperref}

 \journalname{Space Science Reviews}
\begin{document}

\title{Recipes for forming a carbon--rich giant planet}

\titlerunning{Recipes for forming a carbon--rich giant planet}        

\author{Olivier Mousis* \and
        Thibault Cavali\'e \and
        Jonathan I. Lunine \and
        Kathleen E. Mandt \and
        Ricardo Hueso \and
        Artyom Aguichine\and
        Antoine Schneeberger \and
        Tom Benest Couzinou \and
        David H. Atkinson \and
        Vincent Hue \and
        Mark Hofstadter \and
        Udomlerd Srisuchinwong
}
\institute{*Corresponding author. \email{olivier.mousis@lam.fr}
        Olivier Mousis \at
        Aix-Marseille Universit\'e, CNRS, CNES, Institut Origines, LAM, Marseille, France\\
        Institut Universitaire de France (IUF), Paris, France\\
        \and
        Thibault Cavali\'e \at
        Laboratoire d'Astrophysique de Bordeaux, Univ. Bordeaux, CNRS\\
        B18N, all\'ee Geoffroy Saint-Hilaire, 33615 Pessac, France \\
        Tel.: +33-540003271\\
        \email{thibault.cavalie@u-bordeaux.fr}           
        \and
        Jonathan I. Lunine \at
        Cornell University, Department of Astronomy, Ithaca NY, USA\\ \email{jlunine@astro.cornell.edu}
        \and
        Kathleen E. Mandt  \at
        Johns Hopkins University Applied Physics Laboratory, 11100 Johns Hopkins Road, Laurel, MD 20723, USA\\
        \email{Kathleen.Mandt@jhuapl.edu}
        \and
        Ricardo Hueso  \at
        Departamento de F\'isica Aplicada, Escuela de Ingenier\'ia de Bilbao, Universidad del Pa\'is Vasco/Euskal Herriko Unibertsitatea UPV/EHU, Plaza Ingeniero Torres Quevedo, 1, 48013 Bilbao, Spain\\
        \email{ricardo.hueso@ehu.eus}
        \and
        Artyom Aguichine  \at
        Department of Astronomy and Astrophysics, University of California, Santa Cruz, CA USA\\
        \email{aaguichi@ucsc.edu}
        \and
        Antoine Schneeberger \at
        Aix-Marseille Universit\'e, CNRS, CNES, Institut Origines, LAM, Marseille, France\\
        \email{antoine.schneeberger@lam.fr}
        \and
        Tom Benest Couzinou \at
        Aix-Marseille Universit\'e, CNRS, CNES, Institut Origines, LAM, Marseille, France\\
        \email{tom.benest@lam.fr}
        \and           
        David H. Atkinson \at
        Whitman College, Walla Walla, WA 99362, USA\\
        \email{atkinsod@whitman.edu}
        \and 
        Vincent Hue \at
        Aix-Marseille Universit\'e, CNRS, CNES, Institut Origines, LAM, Marseille, France\\
        \email{vincent.hue@lam.fr}
        \and 
        Mark Hofstadter \at
        Jet Propulsion Laboratory, California Institute of Technology, 4800 Oak Grove Dr., 91109, Pasadena, CA, USA\\
        \email{mark.hofstadter@jpl.nasa.gov}
        \and 
        Udomlerd Srisuchinwong \at
        Aix-Marseille Universit\'e, CNRS, CNES, Institut Origines, LAM, Marseille, France\\
        \email{udomlerd.srisuchinwong@etu.univ-amu.fr}
}

\date{Received: date / Accepted: date}

\maketitle

\begin{abstract}
The exploration of carbon-to-oxygen ratios has yielded intriguing insights into the composition of close-in giant exoplanets, giving rise to a distinct classification: carbon-rich planets, characterized by a carbon--to--oxygen ratio $\ge$ 1 in their atmospheres, as opposed to giant planets exhibiting carbon--to--oxygen ratios close to the protosolar value. In contrast, despite numerous space missions dispatched to the outer solar system and the proximity of Jupiter, Saturn, Uranus, and Neptune, our understanding of the carbon-to-oxygen ratio in these  giants remains notably deficient. Determining this ratio is crucial as it serves as a marker linking a planet's volatile composition directly to its formation region within the disk. This article provides an overview of the current understanding of the carbon-to-oxygen ratio in the four gas giants of our solar system and explores why there is yet no definitive dismissal of the possibility that Jupiter, Saturn, Uranus, or Neptune could be considered carbon-rich planets. Additionally, we delve into the three primary formation scenarios proposed in existing literature to account for a bulk carbon-to-oxygen ratio $\ge$ 1 in a giant planet. A significant challenge lies in accurately inferring the bulk carbon-to-oxygen ratio of our solar system's gas giants. Retrieval methods involve integrating in situ measurements from entry probes equipped with mass spectrometers and remote sensing observations conducted at microwave wavelengths by orbiters. However, these methods fall short of fully discerning the deep carbon-to-oxygen abundance in the gas giants due to their limited probing depth, typically within the 10--100 bar range. To complement these direct measurements, indirect determinations rely on understanding the vertical distribution of atmospheric carbon monoxide in conjunction with thermochemical models. These models aid in evaluating the deep oxygen abundance in the gas giants, providing valuable insights into their overall composition.


\end{abstract}

\section{Introduction}
\label{sec1}

Along with hydrogen and helium, carbon (C) and oxygen (O) are the most abundant elements in the Universe, and their abundances are {key to understanding} the chemical evolution of protoplanetary disks and the structure of giant planets. The elemental C/O ratio is a potential tracer that relates the volatile composition of a planet atmosphere directly to the disk region in which it formed \citep{Ob11,Mo12,Ag22,Ho22}. The C/O ratio is estimated to be equal to $\sim$0.55 in the protosolar nebula (PSN) {\citep{Lodders2021}}, and comparisons between this ratio and those inferred in the atmospheres of Jupiter, Saturn, Uranus, and Neptune are used to derive hints on their formation conditions \citep{Mo12,Mo20,Al14,Ag22,Cavalie2023}.

C/O ratios have also been estimated in a fairly large number of close-in giant exoplanets, giving birth to a new category of planets, namely the carbon--rich planets (CRPs), defined by a C/O ratio $\ge$~1 in their atmospheres. The first carbon-rich atmosphere was inferred for the very hot Jupiter WASP-12b \citep{Ma11}, and recent surveys of Hot-Jupiters suggest that many of them could display C/O ratios $\ge$~1 \citep{Ho23}. On the other hand, despite their {proximity to} Earth and the large number of space missions sent to the outer solar system, the status of the C/O ratio in Jupiter, Saturn, Uranus, and Neptune is still poorly defined. In the case of Jupiter, the microwave radiometer (MWR) on board the Juno spacecraft provided a measurement of the water abundance in its equatorial region, and found the deep oxygen abundance to be 2.2$^{+3.9}_{-2.1}$ times the protosolar value, considering the 2$\sigma$ uncertainties \citep{Li20}. { A recent reanalysis of the same data suggests a range of oxygen abundance between 1.4 to 8.3 times the solar value \citep{Li24}. Furthermore,} thermochemical and diffusion models derive a deep O abundance that is 0.3$^{+0.5}_{-0.2}$ times the protosolar value, based on tropospheric CO observations and with assumptions regarding vertical mixing in the troposphere \citep{Cavalie2023}. { This estimate, along with the lower bounds of the Juno oxygen measurements, hints at the possibility of Jupiter being a carbon-rich planet, particularly if the bulk carbon abundance aligns with the determination made by the Galileo probe ($\sim$3--5 times the prosolar value; \cite{wo04})}. The bulk O content in Saturn, Uranus, or Neptune is even more elusive, in absence of direct measurements. So far the only existing indirect determination of the deep O abundance in Saturn is the one provided by Cavali\'e et al. (2024, this issue) from thermochemical modelling and tropospheric CO observations. They find that the deep O abundance is 7--15 times the protosolar value \citep{Lodders2021}. With a C abundance measured to be $\sim$9 times the protosolar value \citep{Mo22}, the resulting C/O ratio in Saturn could be then slightly supersolar. In the cases of Uranus and Neptune, the measurement of CO coupled with thermochemical models allowed the finding of an upper limit for the O abundance, set to 37 and 200 times the protosolar value, respectively \citep{Venot2020}, { while the measured C abundance is $\sim$70--100 times the protosolar value \citep{Lodders2021} in the two planets (see Sec. \ref{sec2.1.3})}. These considerations imply that, overall, the possibility that Jupiter, Saturn, Uranus or Neptune could belong to the family of CRPs remains open.

The aim of this paper is to describe the possible scenarios that could explain the formation of the four giants planets with bulk C/O ratios $\ge$~1 in the protosolar nebula. Section \ref{sec2} details the current status of knowledge about the C/O ratio in the four giants of the solar system. In Sec. \ref{sec3}, we review the three main types of formation scenarios that have been proposed in the literature to explain a bulk C/O ratio $\ge$~1 in a giant planet. Sec. \ref{sec4} is devoted to the measurement methods that could be used to derive the C/O ratio in the four giants of the solar system. Sec. \ref{sec5} is dedicated to discussion and conclusions.

\section{Clues about the existence of carbon-rich giant planets in the solar system}
\label{sec2}

This section summarizes the current status of knowledge about the C/O ratio in the four giants of the solar system. It appears that a supersolar C/O ratio cannot be excluded in any of the four giants if one considers the current data. 


\subsection{Jupiter}
\label{sec2.1.1}

The C/O ratio inside Jupiter is uncertain. The Galileo probe mass spectrometer provided what is generally regarded as a deep carbon-to-hydrogen ratio by measuring the CH$_{4}$ abundance at multiple levels to slightly more than 10 bars pressure. The most recent value by \cite{wo04} takes into account a recalibration based on laboratory data, and ranges between 2.7 and 6 times solar, constant with depth within the error bars. In contrast, the oxygen abundance measured in water (H$_{2}$O) by the Galileo probe is always subsolar but increases with increasing depth down to the final measurement at around 22 bars. Such a depth is well below the theoretical base (5--10 bars) of the water cloud for the measured temperature profile, and has generally been assumed to reflect dynamics in the atmosphere rather than the deep abundance \citep{Sh98,At20}. The microwave radiometer onboard the Juno mission has mapped the brightness temperature of Jupiter over a range of latitudes, { emission angles} and wavelengths. { The fluctuating presence of ammonia (NH$_{3}$) poses challenges in deducing the water abundance due to its significant contribution to microwave opacity. However, in the lowest northern latitudes, ammonia seems to maintain a relatively consistent depth abundance profile. In this region, the deep water abundance (down to 30 bars) has been initially determined to range between 0.1 to 7.5 times the solar value at the 2-$\sigma$ confidence interval \citep{Li20}. A recent reanalysis of these data, which operates under the assumption that Jupiter's equatorial region is super-adiabatic rather than adiabatic, suggests a range of oxygen abundance between 1.4 to 8.3 times the solar value, with the optimal estimate at about 4.9 solar \citep{Li24}.} 

{ Indirect estimates of water abundance also come from the observations of convective eruptions in the planet in energetic moist convective events. Models of convective storms in the planet applied to these observations suggest lower limits of water abundance higher than 1.0 times solar. This abundance is required by the models to generate the expanding rate and high clouds observed on the upper troposphere in specific convective storms in the equatorial region \citep{St86} and south latitudes \citep{In22}, with water abundance close to 2.0 times solar as the preferred value. The cloud tops of the most energetic storms observed in the north temperate belt, arriving to altitudes of 60 mbar, seem to require water abundances of 2--3 times solar \citep{Sa08}. However, \cite{Sa22} limit the water abundance in these storms to values lower than 2 times solar based on the occurrence rate of these convective outbreaks, with this value being compatible with most theoretical works on moist convective storms in the Jovian atmosphere. Interior models also provide some constraint to the water abundance because fits to the gravity field }determined by Juno are sensitive to the tradeoff between the deep temperature profile and the abundance of heavy elements in the bulk interior, provided one knows the equation-of-state of hydrogen and helium \citep{st20}. The equation-of-state is, in fact, not well enough known to avoid a plethora of different interior models, but in general they severely limit the heavy element abundance to between one and two times solar \citep{He22,Ho23}, in disagreement with the Galileo carbon abundance and the Juno-derived NH$_3$ abundance in the presumably well-mixed low northern latitudes. This is puzzling and has led to serious consideration of the idea that the envelope abundance as measured by Galileo is a high-metallicity veneer with a different composition than the bulk interior \citep{sh22}. This idea would be consistent with atmospheric models of moist convection that require abundances of atmospheric water with values near 1.0--2 times solar \citep{In22,Sa22}.

{ One way to resolve this dilemma is to determine the oxygen abundance at the kilobar level through measurements in the upper troposphere of disequilibrium species whose abundances are sensitive to the elemental C/O and vigor of vertical transport. For instance, \cite{Be02}, \cite{Vi10}, and \cite{Wa15} found oxygen abundances that range from subsolar to supersolar values, the main differences arising from the use of different chemical schemes. Such modeling effort relies on the measurement of the abundances of the species that are involved in the chemical equilibrium occurring at the kilobar level. For the C/O ratio, the species involved are H$_2$O, CH$_4$ and CO, where H$_2$O is the unknown of the model. \cite{Cavalie2023} adopted the upper tropospheric abundances of CH$_4$ and CO as measured by Galileo \citep{wo04} and from ground-based observations \citep{Be02,Bj18}}, respectively. With a nominal vertical mixing coefficient of 10$^8$ cm$^2$/s constant with altitude, their analysis yields O values that range from solar to subsolar, i.e.  0.3$^{+0.5}_{-0.2}$ times the protosolar value of \cite{Lodders2021}, implying that C/O could be greater than unity (C/O = 6$^{+10}_{-5}$). {    Alternatively, \cite{Cavalie2023} demonstrated that introducing a radiative layer around the CO quench level results in an oxygen abundance that is approximately consistent with the lower end of the new \cite{Li24} determination, while not violating the constraint on O from interior modeling.

To date, comprehensive analyses of the Juno data, alongside investigations into convective storms and thermochemical models of the envelope, consistently suggest that the oxygen abundance in Jupiter's envelope might be half, or less, the abundance of carbon measured by the Galileo probe \cite{wo04}. This finding hints at the possibility of Jupiter conforming to the criterion of a C/O ratio $\ge$ 1.} Further details are discussed in Cavali\'e et al. (2024, this issue).

\subsection{Saturn}
\label{sec2.1.2}

Unlike Jupiter, only the abundances of a limited number of molecules have been measured in situ in Saturn. The main carbon carrier in the gas giant troposphere and stratosphere is methane. On Jupiter and Saturn, this molecule diffuses from the troposphere up to the stratosphere, where it can be photolyzed to produce a wealth of hydrocarbons (see, \textit{e.g.,} \cite{Moses2005a, Moses2005b, Hue2015, Hue2018}). 
Saturn's methane abundance has been measured from remote sensing observations using ground- and space-based observatories (\textit{e.g.,} \cite{Karkoschka1992, Buriez1981, Kerola1997, Courtin1984, Flasar2005}). Some of the more recent measurements were provided thanks to Cassini-CIRS, from which \cite{Fletcher2009} derived a methane mole fraction of (4.7 $\pm$ 0.2) $\times$ 10$^{-3}$, which corresponds to a carbon enrichment of 10.9 $\pm$ 0.5 over the solar values, assuming protosolar abundances from \cite{Grevesse2007}.

As far as oxygen is concerned, water is the main oxygen carrier in Saturn's atmosphere. Similarly to the methane situation on Uranus and Neptune, water is depleted near the cold tropopause trap. The detected stratospheric water on Saturn therefore results from external sources resulting from rings and satellites interaction (e.g., \cite{Connerney1984, Feuchtgruber1997, Hartogh2011, Cassidy2010, Cavalie2019, Moses2023, Moore2015, ODonoghue2013}), the continuous supply of interplanetary dust particles \citep{Landgraf2002, Poppe2016, Moses2017}, and/or possible cometary impacts \citep{Cavalie2009, Cavalie2010}. Tropospheric water was first detected in Saturn by the ISO spacecraft. Fits of the ISO-SWS spectra suggested a [H$_2$O]/[H$_2$] ratio of $\sim$2 $\times$ 10$^{-7}$ below the 3-bar level \citep{DeGraauw1997}. However, the amount of water measured at that pressure level was found to be undersatured and therefore does not represent the planet’s bulk oxygen content. Similarly to Jupiter, observations of convective storms in Saturn suggest elevated values of the solar water abundance from 1--10 times solar to reconcile the observations of moist convective storms \citep{Hu04,Li15}. As in Jupiter, the values required by models of storms could be representative  of local enhancements of water, and not fully representative of a global average.

On the other hand, the deep O abundance is estimated by Cavali\'e et al. (2024, this issue) from thermochemical modelling and tropospheric CO observations. They find that the deep O abundance is 7--15 times the protosolar value of \cite{Lodders2021}. With a C abundance measured to be $\sim$9 times the protosolar value \citep{Mo22}, the resulting C/O ratio in Saturn could be then slightly supersolar.

\subsection{Uranus and Neptune}
\label{sec2.1.3}
{ Spectroscopic compositional measurements at Uranus and Neptune remain difficult given their distance}, cold tropo\-pause temperatures, as well as small temperature gradient in the stratosphere of Uranus which limits spectral line intensities. Methane condenses in these atmospheres and direct observations in the thermal infrared only probe the depleted stratosphere \citep{Lellouch2015}. More complex techniques using reflective spectroscopy are then used to constrain the tropospheric CH$_4$ abundance. Surprisingly, both ice giants present equatorial regions with about 4\% of CH$_4$ while their polar regions appear to be depleted by a factor of $\sim$2 \citep{Karkoschka2009,Karkoschka2011,Sromovsky2014}. These meridional variabilities are possibly caused by tropospheric circulation (\textit{e.g.,} \cite{Fletcher2020} and references therein). This was recently confirmed by \cite{Irwin2021} for Neptune, though derived abundances were somewhat slightly higher, i.e., 4--6\%~in the equatorial region and 2--4\%~in the southern polar region. Error bars on the deep CH$_4$ abundance in both planets thus remain relatively large. Direct detection of tropospheric water in the atmospheres of Uranus and Neptune has remained elusive given their cold tropopause temperatures. Water vapor is locked below the water cloud level, which likely resides at pressures $>$100 bar \citep{Cavalie2017, At20}. Deriving the deep oxygen in the ice giants thus requires the use of indirect measurements and thermochemical modeling. From the detection of CO in the atmosphere of Neptune \citep{Marten1993} and using estimates of the deep atmospheres vertical mixing, \cite{Lodders1994} found that Neptune could have a rather extreme oxygen enrichment factor of 400 times solar, difficult to explain with formation models (\textit{e.g.,} \cite{mou18}). The lack of direct detection of CO in the atmosphere of Uranus resulted in the derivation of an O/H enrichment upper limit of 260 times solar \citep{Lodders1994}. 

Nearly 30 years later, the situation has slightly evolved. Improvements in the knowledge of the vertical distribution of CO in Uranus and Neptune \citep{Lellouch2005,Teanby2013,Cavalie2014,Teanby2019} coupled with better kinetics have led \cite{Venot2020} to re-evaluate the deep oxygen abundances of the ice giants. They find O/H ratios of 200 and $<$37 times the protosolar value for Neptune and Uranus, respectively, when assuming the equatorial CH$_4$ abundances of \cite{Karkoschka2009,Karkoschka2011,Sromovsky2014} and adopting the solar abundances of \cite{Lodders2021}. As a consequence, Uranus and Neptune present, at face value, a dichotomy regarding their C/O ratios. While Neptune is nominally more enriched in oxygen than in carbon, it is the reverse situation for Uranus, which could be more carbon-rich than oxygen-rich. 
However, and even if unlikely, CO observations are in principle compatible with very low tropospheric CO abundances \citep{Teanby2019}, which would result in a much lower deep oxygen abundance.

\section{Formation scenarios}
\label{sec3}

\begin{figure}[h!]
\centering
\includegraphics[width=0.8\linewidth]{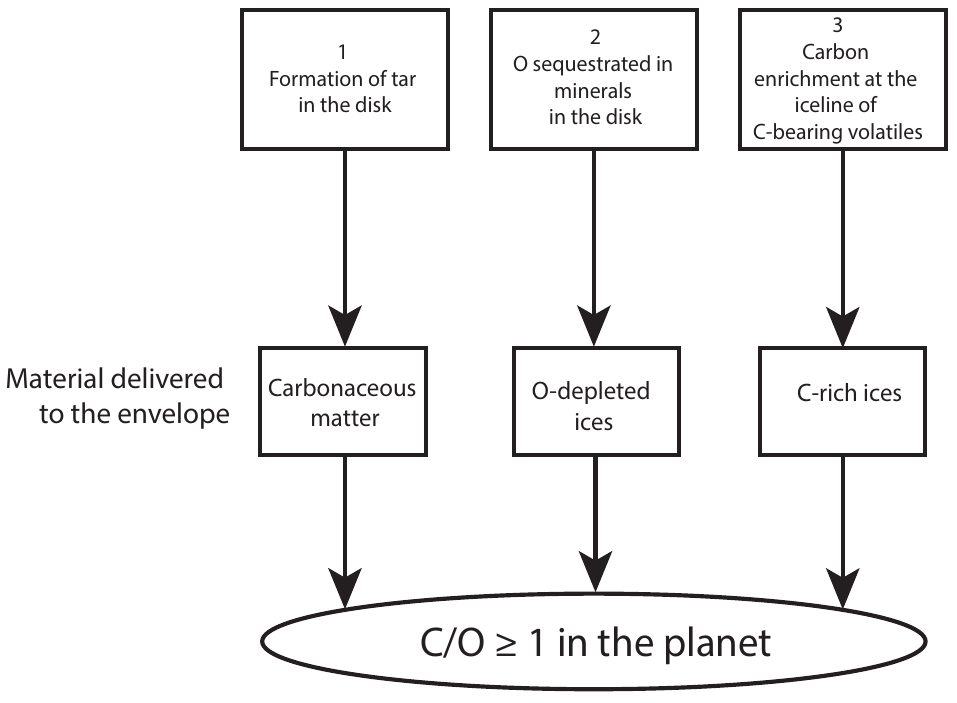}
\caption{A schematic summarizing the different formation scenarios of carbon--rich giant planets. Each scenario explores the accretion of solid matter with a bulk C/O ratio $\ge$1 by the forming giant planet under specific circumstances within the PSN (see text for details).}
\label{fig:scenario}
\end{figure}

Three distinct formation scenarios have been proposed in literature to explain the occurrence of a C/O ratio greater than or equal to 1 in a giant planet. These scenarios include the formation of giant planets from tar, the sequestration of PSN oxygen in refractory matter, and the planetary growth along the ice lines of C-dominating volatiles. { Each of these scenarios is discussed below and visually depicted in Fig. \ref{fig:scenario}.}

\subsection{Tar planets}
\label{sec3.1}

\begin{figure}[h!]
\centering
\includegraphics[width=1\linewidth]{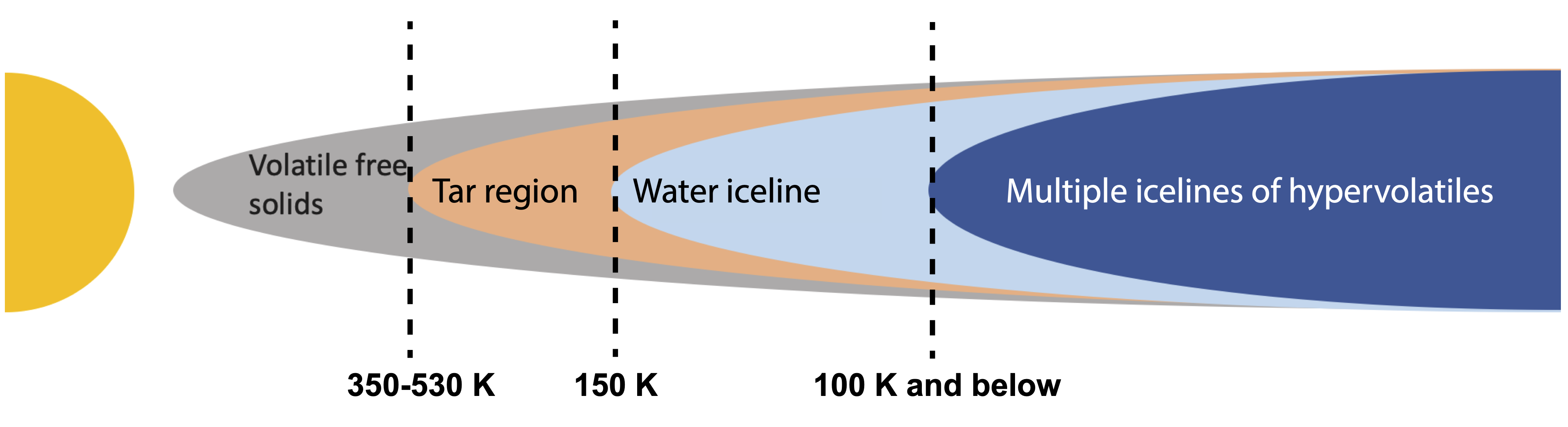}
\caption{A schematic of the relative locations of the different condensation lines in the PSN, with their corresponding temperatures. Carbonaceous material forms in the region between the tar line and the water ice line. Material accreted by giant planets and originating from this region of the PSN should be rich in organics.}
\label{fig:tar}
\end{figure}

One possibility for how carbon-rich planets can form is the scenario where the core of a giant planet forms closer to the Sun or star than the water ice line, but beyond the ``tar line". { This term designates the location where carbonaceous matter undergoes condensation or evaporation, characterized by temperatures ranging between approximately 350 and 530 K within a protoplanetary nebula.} The relative locations of these lines in the protosolar nebula (PSN) are illustrated in Fig. \ref{fig:tar}. This scenario was initially proposed for Jupiter by \cite{lo04}, based on the high C/O ratio measured by the Galileo Probe Mass Spectrometer \citep{wo04}. In the region between the tar line and the water ice line, water is not able to condense, but carbonaceous material can either form by nonequilibrium processes or survive if it originates from the interstellar medium (ISM), thus constituting two possible sources of carbon--rich matter in this region of the PSN \citep{lo04}. Solids quickly grow to large sizes because the solid organics in this region are stickier than at any other location in the disk \citep{Ko02}. The presence of a ``soot line'' located at a somewhat imprecise $\sim$1000--2000 K temperature range has also been argued to account for the high abundance of C$_2$H$_ 2$ observed in protoplanetary disks. The observed C$_2$H$_ 2$ would result from the thermal destruction of PAHs originating from ISM when they cross the soot line \citep{Kr10}.

Although this theory proposes what might be a physically plausible mechanism for the rapid growth of a planetary core that could result in the formation of a giant planet, the application to Jupiter's core formation is not supported by the observations. For example, the heavy noble gases --Ar, Kr, and Xe-- are all enhanced relative to H in Jupiter's atmosphere to a similar degree compared to carbon. When using CI chondrites as an analog for the material that formed Jupiter's core, the heavy noble gases should be depleted relative to solar values \citep{Lodders2021}. { Furthermore, the nitrogen isotopes in Jupiter's atmosphere \citep{Ow01}} are the same as the solar value measured in the solar wind \citep{Marty11}. This requires most of the nitrogen that formed Jupiter to have originated as N$_2$ in the PSN rather than nitrogen in organics that would have been enhanced in the heavy isotope \citep{Ma14}. Although it is possible that Jupiter captured gas that was enhanced in the heavy noble gasses, it would have also captured similar enhancements in nitrogen leading to a much higher N/C ratio than has been measured. Finally, the theoretical impetus for such a { chemically--based} particular mechanism, namely the need to accrete Jupiter's core quickly, appears to have been solved by treating aerodynamic drag of ``pebbles'' in the cm-to-meter size range \citep{Jo17}.

\subsection{Oxygen depletion by sequestration in refractory species}
\label{sec3.2}

{ In this scenario, the presence of a supersolar C/O ratio within the giant planet can be attributed to the devolatilization process of refractory materials within its envelope.} Sequestration of oxygen in rock-- and metal--forming species (those involving Mg, Si, Fe, Al,  etc.) must be considered when the carbon-to-oxygen ratio exceeds the solar value of C/O =  0.55 { in the PSN}. Even here, the abundances of the rock forming elements are sufficient to depress the amount of oxygen available to form water by a noticeable amount if the dominant carbon--bearing molecule in the disk is CO. When the C/O ratio exceeds 0.8 { in the initial gas phase of the PSN}, the depletion of oxygen is severe, such that it is bound up in rock, metal oxides, CO and CO$_2$ to the exclusion of water \citep{Pe19}. The same is not true if the disk chemistry is reducing enough that CH$_4$ dominates (although the kinetics of maintaining a CH$_4$-rich disk are difficult; see \cite{Pr89}). The presence of refractory organics also decreases the dependence of the water abundance on the C/O ratio \citep{Wo08}.  For our own solar system, recent work suggests that the presence of refractories in the protoplanetary disk could bias the apparent C/O ratios in the giant planets upward by as much as 1/3 \citep{Fo23}, but is still not enough to explain C/O values exceeding unity.  
 \subsection{Role of icelines in the protosolar nebula}
\label{sec3.3}

\begin{figure}[h!]
\centering
\includegraphics[width=1\linewidth]{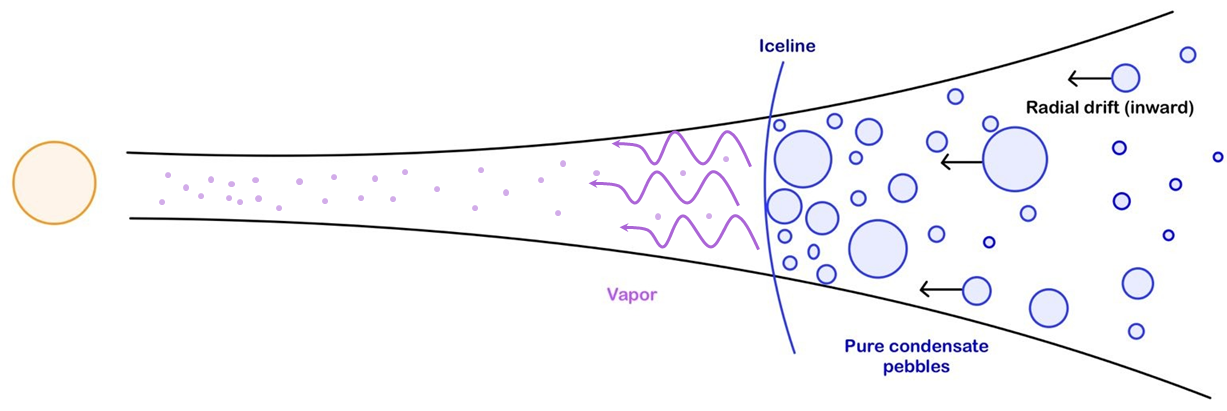}
\caption{A schematic of the overall system under consideration: the protosun surrounded by the PSN. The radial distribution of the volatile species under consideration is affected  by the diffusion of its vapor out past the condensation front, where the vapor condenses into icy particles, and the radial drift back inward of those particles.}
\label{fig:cycling}
\end{figure}

An iceline is defined as the distance where the surface density of vapor of a given volatile species is equal to that of its solid form in the PSN. Inside the snowline, water ice evaporates into water vapor. Outside the snowline, ice is present due to the condensation of vapor, though the motion of particles within the disk allows for solids to exist in front of this line as well as some vapors to exist beyond, due to the kinetics of condensation/sublimation. These ice lines are locations where condensation/vaporization cycles can enhance their abundances in both solid and vapor forms,  
\citep{St88,Cy98,Al14,Mo20,Mo21,Ag22,sch23}. When icy pebbles cross the iceline of a given species moving inward towards the Sun, its solid phase vaporizes and the released vapor diffuses both inward and outward, implying that a given fraction of the vapor crosses the iceline again. This fraction of vapor recondenses back onto icy grains which cross the iceline, and feed again its interior region with vapor. A key point is that the outward diffusion of vapor is shown to be faster than its replenishment inside the icelines by sublimating ices. This leads to depletion in vapors inside the icelines and a concentration of solids at the iceline positions. By doing so, this cycle produces an enrichment peak of the species at the iceline which can reach or exceed $\sim$10 times the initial abundance in icy grains \citep{Ag22,sch23}. A schematic of this mechanism is shown in Fig. \ref{fig:cycling}. This recycling process applies to all oxygen-- and carbon-bearing volatile species, including H$_2$O, CO, CO$_2$ and CH$_4$, and can shape the C/O ratio in the material accreted by gas and ice giants during their growth \citep{Mo19,Mo20,Mo21,Ag22,Cavalie2023}.

Pure condensates, and perhaps clathrates, are the two crystalline structures that can form during the cooling of the outer PSN \citep{sch23}.  Clathrates are believed to exist in large amounts in the crust of dwarf planets and the satellites of outer planets such as Titan and Pluto \citep{Mo02,To06,Bo19,Ka19}. The potential presence of clathrates in the PSN depends on the local availability of crystalline water, as well as on their kinetics of formation at low-pressure and low-temperature conditions \citep{lun85,gau05,Mo10}. The formation of Comet 67P/churyumov-Gerasimenko from a mixture of clathrates originating from the PSN has been invoked to explain the argon deficiency and the presence of ultravolatiles observed in the interior of this body by the Rosetta spacecraft \citep{Mo16,Lu16}.
 
\begin{figure}[h!]
\centering
\includegraphics[width=0.9\linewidth]{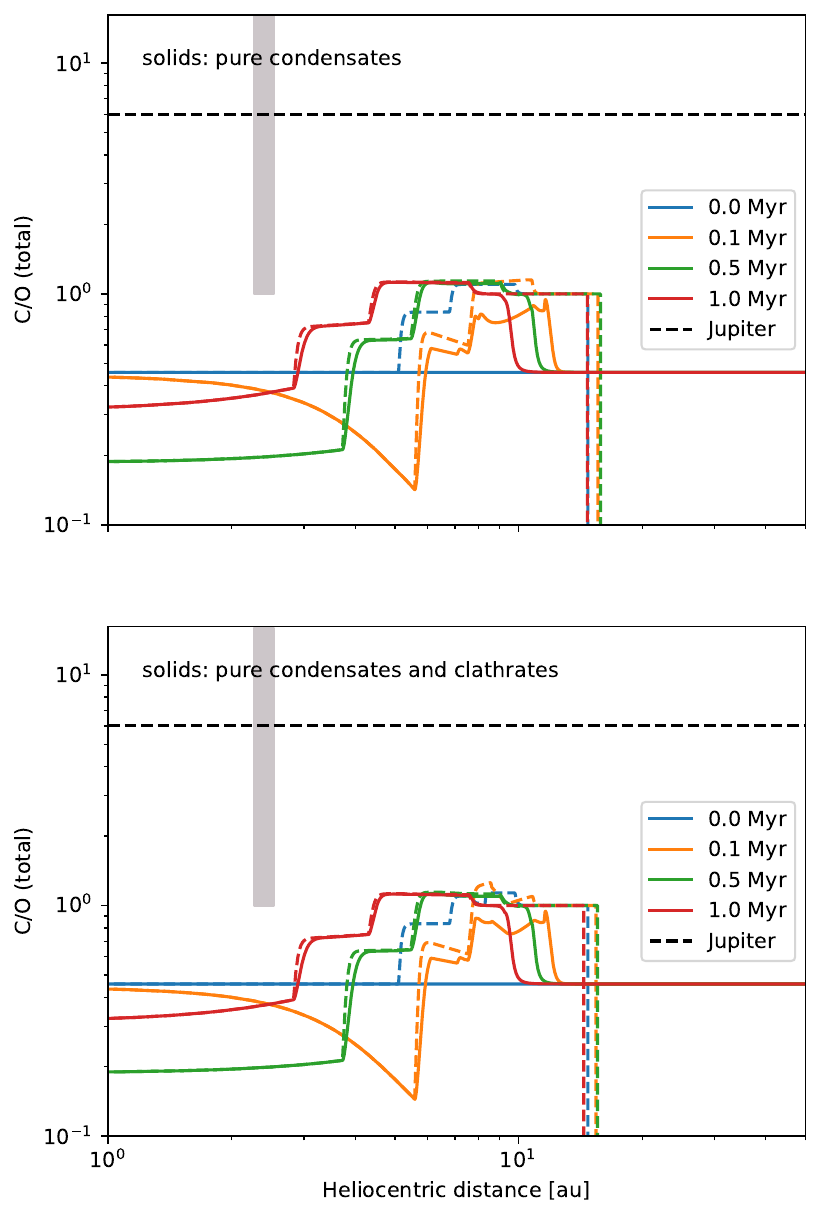}
\caption{C/O ratio in vapor phase (dashed lines) and in both vapor and solid phases (solid lines) as a function of heliocentric distance after 0, 0.1, 0.5, and 1 Myr of PSN evolution. Two scenarios are considered: solids composed of pure condensates (top panel) and of a mixture of pure condensates and clathrates (bottom panel). The vertical grey rectangle and the black dashed horizontal line correspond to the 1--$\sigma$ error bar and the nominal value associated with the C/O ratio estimated in Jupiter, respectively \citep{Cavalie2023}. The blue horizontal line corresponds to the protosolar C/O ratio.}
\label{fig:C-O_ratio}
\end{figure}

{ Figure \ref{fig:C-O_ratio} illustrates the C/O abundance profile as a function of heliocentric distance in the PSN following 0, 0.1, 0.5, and 1.0 Myr of disk evolution, utilizing the methodology outlined in \cite{sch23}. The model tracks the evolution of volatile species in both vapor and ice phases, including species trapped in clathrate hydrates. It incorporates the advection-diffusion of vapor species and considers the dynamics of dust and pebbles using the two-population model from \cite{Bi12}. 
Two scenarios are explored: the formation of pure condensates alone in the PSN and the concurrent formation of pure condensates and clathrates. In both scenarios, C and O are distributed among H$_2$O, CO, CO$_2$, and CH$_4$, with  protosolar abundances drawn from \cite{Lo09}, resulting in a C/O ratio of approximately 0.46. The initial mixture assumes a volume ratio of CO:CO$_2$:CH$_4$~=~10:4:1 (see \cite{sch23} for details). Additionally, for comparison, the figure represents the C/O ratio (6$^{+10}_{-5}$) estimated by \cite{Cavalie2023} within Jupiter's interior.

Figure \ref{fig:C-O_ratio} demonstrates that both icelines and clathration lines of carbon-bearing compounds significantly enhance the C/O ratio, exceeding the protosolar value (C/O $\ge$~1) within the 5--15 AU region between different icelines. If Jupiter formed in this region during the first Myr of PSN evolution, it could have accreted C-rich solids, potentially resulting in a C/O ratio consistent with the estimate of \cite{Cavalie2023}. However, one should note that this study offers insight into the composition of the PSN before the formation of planetary cores, without modeling the formation of planetesimals and their subsequent evolution via pebble accretion and gas-runaway accretion \citep{He14,Bi15}.

We limit our modeling to 1 Myr of PSN evolution, as the disk thermodynamic profiles reach steady states after this period, with insignificant changes thereafter. Moreover, the timescale of planet formation, which remains a subject of debate, could be shorter than 1 Myr, according to \cite{la12,He14,la14}}.


\section{Measurements needed to assess the C/O status in giant planets}
\label{sec4}


To determine the ratio of carbon-to-oxygen in the atmospheres of the giant planets requires measurements of the deep (well-mixed) abundances of both oxygen and carbon. Although carbon does not condense in the form of methane (CH$_4$) in the warmer atmospheres of Jupiter and Saturn and is therefore assumed to be well-mixed throughout the atmosphere of these planets, the situation is more complex at the ice giants where the much colder temperatures result in methane condensing into clouds in the upper troposphere. Oxygen is primarily contained within water which condenses into clouds in the troposphere of all the giant planets. { According to equilibrium cloud condensation models (ECCMs) \citep{We73,At20}}, the base of the water clouds is expected to be at about 5--8 bars at Jupiter depending on the abundance of oxygen and possible mixing with NH$_3$ and H$_2$S, at about 10--20 bars at Saturn, and much deeper at the ice giants, at pressures of 50–-200 bars at Uranus and 40-–500 bars at Neptune.

{ Obtaining in situ measurements to determine the global, deep, and well-mixed abundance of water presents considerable challenges. These challenges are due to the depth of water clouds based on the expected atmospheric temperature profiles nd great range of potential water abundances, and ECCMs of the atmosphere. Achieving a representative measure of the giant planet's water abundance necessitates the deployment of a probe capable of descending well below the water cloud condensation level to access the well-mixed region of the atmosphere.}
Although the Galileo probe, despite returning data down to 22 bars on Jupiter, descended through a region devoid of clouds called a 5--micron hotspot and hence did not reach the well-mixed water region \citep{Or98}, the base of the 5-bar water cloud base predicted by ECCMs is well within the reach of descent probes. At Saturn, the temperature at a given pressure is lower than at Jupiter and the water cloud base is expected to be between 10--20 bars, still well within reach of an atmospheric descent probe mission. However, the ice giant water clouds are expected to be significantly deeper and far beyond the depth probes can reach with current technologies. To reach the well-mixed region beneath the water clouds at Uranus and Neptune would require a probe to survive and return data from levels potentially as deep as several 100's of bars. 
{ To ensure the survival of a descent probe at depths exceeding 100 bars where both pressure and temperatures will be extreme, several critical challenges must be addressed. These include developing power systems sufficient to sustain the probe throughout its descent and stay operational under such conditions. Moreover, crafting a robust telecommunications system capable of penetrating the dense and highly absorbing atmosphere above the probe will not be trivial. Meeting these challenges demands groundbreaking advancements in engineering and technology.}


The only technique currently available by which measurements can be made of the deep abundances of water and other key constituents of the giant planets is by measurement of microwaves emitted by the deep atmosphere over a range of frequencies from UHF ($\sim$300MHz) to Ka-band ($\sim$30 GHz), corresponding to wavelengths of about 100 cm to 1 cm, respectively \citep{dePater2023}. Microwave emissions from giant planet atmospheres originate from the tropopause and the lowest frequencies (longer wavelengths) emanate from the deepest atmosphere, possibly 1000 bars, much deeper than current probes are designed to operate. The microwave spectrum emitted by giant planet atmospheres depends on the temperature and composition of the atmosphere. Measurement of the microwave spectrum at multiple frequencies provides a means by which abundances of key microwave opacity sources can be determined. To accurately measure the abundance of key constituents in the deep atmosphere therefore requires an accurate model of the thermal structure of the atmosphere \citep{Lu13,Le17}.  

Models of the thermal structure of the deep atmosphere are primarily based on physical principles, but can also be strongly constrained by measurements of temperatures at a single location in the upper atmosphere provided by in situ measurements from an atmospheric descent probe \citep{Se98}. Although measurements of the microwave emissions from the giant planets can be made from Earth, there are limitations. First, synchrotron radiation due to extragalactic sources as well as from Jupiter provide enough radio noise that Earth-based microwave measurements are either impossible, or at very low SNR (e.g. \cite{Co15} for Saturn). Additionally, due to the great distance to the outer solar system, the spatial resolution of outer planets possible { at these wavelengths from Earth is quite poor}. { Although large baseline interferometers such as the Atacama Large Millimeter Array (ALMA), or the Very Large Array (VLA) can peer into the deep atmosphere of Uranus and Neptune resolving bands and deep structures \citep{To19,Mol21}, the deepest layers that can be sensed are about 50 bar, well above the layers where emissions from water should be observable. }Therefore, the only reasonable means by which microwave measurements of the giant planets can be made is remote sensing from flyby or orbiting spacecraft. The Juno orbiter currently flies in a polar orbit that passed beneath Jupiter’s radiation belts to make microwave abundance measurements of H$_2$O and NH$_3$ to 100 bars or deeper, less contaminated by the noise from Jupiter’s intense synchrotron radiation { \citep{Bo17,Janssen2017,Li20,Li24}}. 

To complete the measurement of the C/O ratio in giant planet atmospheres requires in situ measurements of atmospheric carbon by a mass spectrometer carried by an atmospheric descent probe. At the gas giants Jupiter and Saturn, carbon in the form of methane does not condense and is therefore well-mixed throughout the atmosphere. { An atmospheric probe at any location on the gas giants} would measure the well-mixed abundance of methane from which a value for the global inventory of carbon can be retrieved. The situation is more difficult at the ice giants Uranus and Neptune where temperatures are low enough that carbon condenses in the form of CH$_4$ clouds { at levels of 1--2 bars \citep{Mo22}.} If ECCMs are correct, a relatively shallow ice giant atmospheric entry probe would pass through the methane clouds into the region beneath the clouds where measurements of atmospheric methane by a probe mass spectrometer would provide the global atmospheric carbon content \citep{Atk20}. 

When the results of the measurements of NH$_3$ at Jupiter made by Juno are considered along with the fact that CH$_4$ condenses in ice giant atmospheres, it is questionable as to whether simple ECCMs and abundance measurements of CH$_4$ abundances made in situ by probes can truly reflect the composition of the bulk atmosphere. { On the other hand, in the case of Juno at Jupiter, it is plausible that the observed NH$_3$ distribution is influenced by the presence of the water cloud, where moist convection establishes unexpected compositional gradients well below the cloud base \citep{Su14,Le17}. However, for the ice giants, CH$_4$ itself is abundant enough to inhibit convection \citep{Gu95}. Consequently, the vertical displacement between the well-mixed mixing ratio and the condensation level might be primarily governed by the inhibiting species (H$_2$O on Jupiter, CH$_4$ on Uranus), thereby influencing the distribution of trace species (NH$_3$ in Jupiter and CH$_4$ in Uranus). In this scenario, the depth difference of the well-mixed tracer layer in Uranus and Neptune may not be as significant. It should be also noted that, alongside CH$_4$, a shallow probe outfitted with a sufficiently sensitive and high-resolution mass spectrometer could detect the abundance of CO \citep{Cavalie2020}. 

Measurements of CO an thermochemical models would provide diagnostic assessments of the oxygen distribution in the deep atmospheres of the giant planets, complementing determinations made through microwave observations.}



\section{Conclusions}
\label{sec5}

In this review, we provide an overview of the ongoing debate surrounding the carbon-to-oxygen (C/O) ratio in the bulk interiors of the four gas giants in our solar system, suggesting the possibility that some may indeed be carbon-rich. Various formation mechanisms have been proposed in the literature to explain this potential scenario. For instance, carbon-rich giant planets could have originated if their cores formed closer to the host star, within the region bounded by the tar line and the water ice line \citep{lo04}. However, this hypothesis doesn't apply to Jupiter, as the supersolar abundances of noble gases observed in its atmosphere are inconsistent with the notion that they were delivered alongside carbon-rich refractory matter.  { As of now, no compelling mechanism has been put forward to explain the supersolar abundances of the heavy noble gases within the framework of this hypothesis.} Another scenario suggests that a significant portion of oxygen was sequestered in rock-- and metal--forming species within the protosolar nebula, rather than forming water \citep{Fo23,Pe19}. However, this would necessitate assuming an initial C/O ratio in the protosolar nebula gas much higher than the solar value (0.55) before the formation of the first solids. Intriguingly, this hypothesis finds support in measurements of the density and moment of inertia of icy moons and dwarf planets, which indicate accretion from low-density carbonaceous components in their rocky cores \citep{Re23}, suggesting that condensed organic carbon could be a significant constituent of outer solar system planetary bodies. Another proposed mechanism involves the growth of gas giants near the icelines of key carbon-bearing volatiles (such as CH$_4$, CO, CO$_2$, etc.) \citep{St88,Cy98,Al14,Mo20,Mo21,Ag22,sch23}. In these regions, giant planets could have naturally accreted a mixture of ices and vapors with C/O ratios equal to or greater than 1.

{ Despite encountering challenges in the case of Jupiter, the notion of giant planet formation from tar requires consideration. Moreover, it is worth highlighting that both of the latter scenarios may coexist within the PSN. The constitution of Jupiter's envelope could plausibly emerge from the devolatilization of carbonaceous material, alongside the vaporization of ices that formed in proximity to the CO and CO$_2$ icelines, among other possibilities. In the absence of observational constraints indicating challenges akin to those observed in Jupiter, the three scenarios under discussion could conceivably operate simultaneously in other planetary systems to explain the existence of carbon-rich giant planets. If planetary embryo migration spans a broad distance range within a protoplanetary disk, each forming giant planet could accrete material formed through each of the three distinct scenarios. These scenarios would operate at various distance ranges within the disk, allowing for the accumulation of diverse solid materials.}

The primary challenge persists in accurately determining the bulk carbon-to-oxygen (C/O) ratio of the gas giants in our solar system. Current retrieval methods involve a combination of in-situ measurements with mass spectrometers aboard entry probes and remote sensing observations using microwave wavelengths from orbiters. However, these retrieval methods do not guarantee the precise derivation of the deep C/O abundance within the giant planets, as they are constrained by probing depth limitations, typically within the range of 10 to 100 bars. Thus far, only indirect determinations, reliant on understanding the vertical distribution of CO in the giant planets alongside enhanced thermochemical models, can aid in assessing the deep oxygen abundances. { This underscores the critical necessity for probe mass spectrometers and/or tunable laser spectrometers to possess sensitivity not only sufficient to measure elemental main carriers (e.g., CH$_4$, H$_2$O, NH$_3$, etc.) but also various less abundant species \citep{Cavalie2020,We23}.}

\begin{acknowledgements}
OM and TC acknowledge support from CNES. The project leading to this publication has received funding from the Excellence Initiative of Aix-Marseille Universit\'e--A*Midex, a French ``Investissements d’Avenir program'' AMX-21-IET-018. This research holds as part of the project FACOM (ANR-22-CE49-0005-01\_ACT) and has benefited from a funding provided by l'Agence Nationale de la Recherche (ANR) under the Generic Call for Proposals 2022. RH was supported by Grant PID2019-109467GB-I00 funded by MCIN/AEI/10.13039/501100011033/ and by Grupos Gobierno Vasco IT1742-22. VH acknowledges support from the French government under the France 2030 investment plan, as part of the Initiative d’Excellence d’Aix-Marseille Université – A*MIDEX AMX-22-CPJ-04. The work of MH was performed at the Jet Propulsion Laboratory, California Institute of Technology, under contract with NASA. JL was supported by the Distinguished Visiting Scientist Program at JPL. This material is based upon work supported by NASA'S Interdisciplinary Consortia for Astrobiology Research (NNH19ZDA001N-ICAR) under award number 19-ICAR19$\_$2-0041.
\end{acknowledgements}


\bibliographystyle{spbasic}      
\bibliography{biblio}   

%
%

\end{document}